# Enabling On-Demand Database Computing with MIT SuperCloud Database Management System

Andrew Prout, Jeremy Kepner, Peter Michaleas, William Arcand, David Bestor, Bill Bergeron, Chansup Byun, Lauren Edwards, Vijay Gadepally, Matthew Hubbell, Julie Mullen, Antonio Rosa, Charles Yee, Albert Reuther

MIT Lincoln Laboratory, Lexington, MA, U.S.A.

*Abstract*—The MIT SuperCloud database management system allows for rapid creation and flexible execution of a variety of the latest scientific databases, including Apache Accumulo and SciDB. It is designed to permit these databases to run on a High Performance Computing Cluster (HPCC) platform as seamlessly as any other HPCC job. It ensures the seamless migration of the databases to the resources assigned by the HPCC scheduler and centralized storage of the database files when not running. It also permits snapshotting of databases to allow researchers to experiment and push the limits of the technology without concerns for data or productivity loss if the database becomes unstable.

## I. Introduction

The ability to collect and analyze large amounts of data is a growing problem amongst the scientific community. The growing gap between data and users calls for innovative tools that address the challenges faced by big data volume, velocity and variety. There have been many recent advances in database technology trying to solve these problems. However, these new database technologies have come with their own challenges around setup, management and allocating computing resources.

Relational or SQL (Structured Query Language) databases [Codd 1970, Stonebraker et al 1976] have been the de facto interface to databases since the 1980s and are the bedrock of electronic transactions around the world. More recently, key-value stores (NoSQL databases) [Chang et al 2008] have been developed for representing large sparse tables to aid in the analysis of data for Internet search. As a result, the majority of the data on the Internet is now analyzed using key-value stores [DeCandia et al 2007, Lakshman & Malik 2010, George 2011]. In response to the same challenges, the relational database community has developed a new class of array store (NewSQL) databases [Stonebraker et al 2005, Kallman et al 2008, Lamb et al 2012, Stonebraker & Weisberg 2013] to provide the features of relational databases while also scaling to very large data sets.

However, these databases traditionally take significant effort to set up and maintain properly. Once configured on a particular system, it is difficult to move them to another system. This leads to dedicated hardware that goes unused when the database is not in use. While virtual machines could be used to partially address these issues, they introduce additional complexities and performance penalties of their own.

Our MIT SuperCloud supports all three classes of databases: Postgresql (SQL), Accumulo (NoSQL), and SciDB (NewSQL). In addition, a common interface is provided to these different databases via the D4M programming environment [Kepner et al 2012, Kepner & Gadepally 2014]. The MIT SuperCloud database management system allows researchers a high degree of control and flexibility in interacting with the databases without low-level system access or in-depth knowledge of the database configuration. With this system we are able to automate the creation of the databases on our central file system, the launching of these databases onto computing resources in the HPCC; the orderly shutdown, cleanup and state preservation when they are not needed; and the ability to checkpoint and restore a known-good state.

The organization of the rest of this paper is as follows. Section II introduces Accumulo, SciDB, D4M, and the MIT SuperCloud system used to conduct the performance measurements. Section III describes the operational model and user experience of this system. Section IV describes the steps taken to ensure the databases and their data are kept secure from unintended access. Section V shows the performance results and overhead of the system. Section VI summarizes the results.

## II. Technologies

A variety of technologies were used to implement this system. Together, these technologies make up the MIT SuperCloud [Reuther 2013] (see Figure 1). The MIT SuperCloud allows big data applications such as Accumulo and SciDB to run on a supercomputer system.

### A. Accumulo Database

Accumulo is a key-value store where each entry consists of a seven-tuple. Most of the concepts of Accumulo can be understood by reducing this seven-tuple into a triple consisting of a row, column, and value. Each triple describes a point in a table. Only the non-empty entries are stored in each row, so the table can have an unlimited number of rows and columns and be extremely sparse, which makes Accumulo well-suited for storing graphs.

Accumulo is designed to run on large clusters of computing hardware where each node in the cluster has its own data storage. Accumulo uses the Hadoop Distributed File System (HDFS) to organize the storage on the nodes into a single, large, redundant file system. A table in Accumulo is broken up into tablets where each tablet contains a continuous block of rows. The row values marking the boundaries between tablets are called *splits*. A table can be broken up into many tablets, and these tablets are then stored in HDFS across the cluster. Good performance is achieved when the data and the operations are spread evenly across the cluster. The selection of good splits is key to achieving this goal.

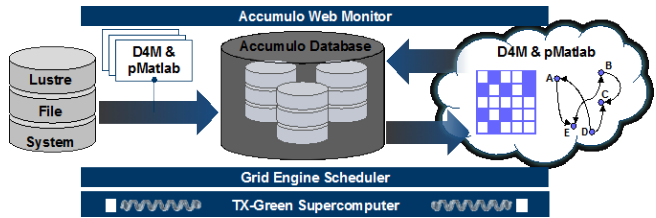

Figure 1. MIT SuperCloud architecture consists of seven components. (1) Lustre parallel file system for high performance file I/O, (2) D4M & pMatlab ingest processes, (3) Accumulo parallel database, (4) D4M & pMatlab analytic processes, (5) Accumulo web monitor page, (6) Grid Engine scheduler for allocating processes to hardware, and (7) the TX-Green supercomputer.

The various Accumulo processes are managed by Zookeeper (zookeeper.apache.org), which is a centralized service for maintaining configuration and naming information, along with providing distributed synchronization and group services.

### B. SciDB

SciDB [Balazinska et al 2009] is a newSQL database that is designed to run on clusters of computing hardware where each node in the cluster has its own data storage. It uses the disks in the node directly for storage of its data. It also uses PostgreSQL for its System Catalog for storage of metadata, such as table definitions and installed plug-ins. An operational PostgreSQL database is a prerequisite to running SciDB.

### C. Lustre parallel file system

The MIT SuperCloud has two forms of storage: distributed and central. Distributed storage exists on the compute nodes that are used for running Hadoop and Accumulo applications. Central storage is implemented using the open source Lustre parallel file system (lustre.org) on a commercial storage array. Lustre provides high performance data access to all the compute nodes, while maintaining the appearance of a single filesystems to the user. The Lustre filesystem is used in most of the largest supercomputers in the world.

The MIT SuperCloud leverages both types of storage to dynamically start, stop, checkpoint, relocate, and restart Accumulo databases by storing their data in the Lustre filesystem when they are stopped and on distributed storage while they are running. This dynamic database management system allows many more Accumulo databases to be hosted on the system than would otherwise be possible. Groups of users can quickly create their own Accumulo databases to share data amongst themselves without interfering with other groups. In addition, because all the Accumulo instances are running directly on the compute nodes, they can run at maximum performance.

### D. Grid Engine scheduler

Supercomputers require efficient mechanisms for rapidly identifying available computing resources, allocating those resources to programs, and launching the programs on the allocated resources. The open source Grid Engine software (https://arc.liv.ac.uk/trac/SGE) provides these services and is independent of programming language (C, Fortran, Java, Matlab, etc) or parallel programming model (message passing, distributed arrays, threads, map/reduce, etc).

The Grid Engine scheduler coordinates the starting and stopping of database instances in the MIT SuperCloud. A user authenticates using a web page that shows them only the databases they are allowed to access. They can then start and stop any of these databases. When a database is started, the MIT SuperCloud database management system communicates to Grid Engine the computing requirements of the database. Grid Engine then finds the computing resources and allocates them to the database, the database files are copied to the appropriate computing nodes, dynamic domain name entries are assigned to the compute nodes, and the database processes are started.

### E. TX-Green hardware

The TX-Green supercomputer consists of 270 HP servers connected to a single Arista core switch using 10 GigE. The Lustre central storage system uses a 1 Petabyte DDN storage array that is directly connected to the core switch using multiple 40 GigE connections. This architecture provides high bandwidth to all the nodes and the central storage. Each server has 32 cores (x86 instruction set), 128 Gigabytes of memory, and 12 Terabytes of storage. The storage is hot-swappable RAID5 so that each node can tolerate one drive failure.

TX-Green is housed in an HP EcoPOD mobile data center that uses ambient air cooling to maximize energy efficiency. The EcoPOD is located near a hydroelectric dam that delivers clean energy that does not contribute green house gases to the environment.

The MIT SuperCloud software stack, which contains all the systems and applications software, resides on every node. Hosting the application software on each node accelerates the launch of large applications (such as databases) and minimizes their dependency on the central storage.

### F. Dynamic DNS

The MIT SuperCloud Dynamic DNS component is used to direct applications to the current resources assigned to run the requested service. Applications, such as databases, can register their IP address upon startup to a well-known name. This well-known name can persist across independent executions of the application assigned to run on different hardware by the Grid Engine scheduler. This allows the well-known name to be used in configuration files, or even stored into data files, without the concern that it will become outdated or need to be manually changed.

The service is based on MyDNS (http://mydns.bboy.net/) which uses a MySQL database to store DNS records. The DNS records are controlled via a simple web service interface that allows records to be created, updated or deleted. The dynamic DNS zone is registered into the network's normal DNS server as a sub-zone so that it is available from both within the cluster and users' workstations. A low time-to-live (TTL) is used on all records in the dynamic DNS zone to ensure that caching DNS servers do not return stale answers for extended periods of time.

*G. D4M analytics library*

D4M is open source software that provides a convenient mathematical representation of the kinds of data that are routinely stored in spreadsheets and large triple store database. Associations between multidimensional entities (tuples) using string keys and string values can be stored in data structures called associative arrays.

Associative arrays can represent complex relationships in either a sparse matrix or a graph form. Thus, associative arrays are a natural data structure for performing both matrix and graph algorithms. Such algorithms are the foundation of many complex database operations across a wide range of fields [Kepner 2011].

Constructing complex composable query operations can be expressed using simple array indexing of the associative array keys and values, which themselves return associative arrays. The composability of associative arrays stems from the ability to define fundamental mathematical operations whose results are also associative arrays. Measurements using D4M indicate these algorithms can be implemented with a tenfold decrease in coding effort when compared to standard approaches [Kepner et al 2012].

III. OPERATIONAL MODEL AND USER EXPERIANCE

The primary goal of the MIT SuperCloud database management system was to create an operational model for these new database technologies that met the needs of the researchers that would use the databases. There are 5 main operations possible in this system: creation, start, stop, create checkpoint, and restore checkpoint. All of these operations take security into consideration in unobtrusive ways. The users primary tool for interacting with the database management system is the Database Status web portal show in Figure 2.

| Folder Name | Type | Status | Actions | | |
|---|---|---|---|---|---|
| classdb01 | Accumulo v1.4.1 | starting | View Info | | |
| classdb02 | Accumulo v1.5.0 | started | View Info | Stop | |
| classdb03 | Accumulo v1.6.0 | stopped | View Info | Start | Checkpoint |
| scidb01 | SciDB 14.3 | stopped | | Start | Checkpoint |
| scidb02 | SciDB 14.3 | started | | Stop | |

Figure 2. The MIT SuperCloud database status web page. The **Folder Name** column represents the name of the directory on central storage that houses the database properties and the data files when not running. The **Type** column represents the type of database, Accumulo or SciDB, and version in this example. The **Status** column represents the current status of the database: stopped, starting, started, stopping, or checkpointing. The **Actions** column contains command buttons appropriate to the current status.

*A. Database creation*

Database creation is performed by MIT SuperCloud system administrators. Working with the users, the administrator determines the database type and version, the number of compute nodes needed to run the database, the project's Linux security group, and an appropriate name for the database. The administrator then runs the db_create command.

This command creates a new folder for the database on the Lustre central storage and initializes the database. In the case of Accumulo, this means it also initializes the Hadoop HDFS filesystem. An example of the command is shown in Figure 3.

```
db_create accumulo --num-nodes 4 dbname01 secgroup
```

Figure 3. A sample database creation command. The example shown would create a four node database with the name "dbname01" and permit members of the Linux security group "secgroup" to access the database.

*B. Database startup*

Database startup can be requested by any user in the database's assigned Linux security group. The user requests the database start either through the **Start** button on the database status webpage shown in Figure 2 or by executing the db_start command at the command line. The db_start script performs several checks, such as ensuring the current database status is stopped, and schedules a Grid Engine job to run the database. The Grid Engine job is requested with the "now" option so that it will not queue for later and return an immediate status to the user if there is an error, such as insufficient resources available.

The Grid Engine job is started in a special "db" queue that is configured with prolog and epilog scripts that execute on each node assigned to the job. The prolog script performs all the actions necessary to start the database, including registering the IP address of the assigned node in dynamic DNS, copying the data from central storage to local disk, starting the PostgreSQL, Kerberos, Hadoop, Zookeeper, SciDB and Accumulo services as appropriate for the database type and updating the database status. The actual "job" that runs after the prolog script is complete is just a placeholder that sleeps forever until the database is stopped. The prolog script is not interrupted by a Grid Engine job cancellation request, unlike the job itself, which ensures the database startup operation is not accidentally interrupted leaving the database in an inconsistent state.

*C. Database stop*

Database shutdown can be requested by any user in the database's assigned Linux security group. The user requests the database stop either through the **Stop** button on the database status webpage shown in Figure 2, by executing the db_stop command at the command line, or by issuing a Grid Engine job cancellation on the database's job. Only the user that originally started the database, or a system administrator, can issue a Grid Engine job cancellation request, however. The db_stop script performs several checks, such as ensuring the current database status is started. It then signals for the Grid Engine job to end its infinite sleep.

When the Grid Engine job completes, the epilog script of the "db" queue is executed. This epilog scripts performs all the actions necessary to shut down the database, including stopping all the daemons, copying the data from local disk back to central storage, and de-registering the dynamic DNS. The epilog script is run no matter how the job was ended and is not interrupted by a Grid Engine job cancellation request, which ensures the database shutdown operation is not accidentally interrupted leaving the database in an inconsistent state.

*D. Database checkpoint creation*

Database checkpoint creation can be requested by any user in the database's assigned Linux security group. It can only be requested when the database is not running. The user requests the database checkpoint either through the **Checkpoint** button on the database status webpage shown in Figure 2 or by executing the db_checkpoint command at the command line. The db_checkpoint script performs several checks, such as ensuring the current database status is stopped, and schedules a Grid Engine job to checkpoint the database. The checkpoint Grid Engine job runs only on a single node, even if the database is configured to normally run on multiple nodes. The Grid Engine job is requested with the "now" option so that it will not queue for later and return an immediate status to the user if there is an error, such as insufficient available resources.

*E. Database checkpoint restoration*

Database checkpoint restoration must be performed by a MIT SuperCloud system administrator. It involves removing the current ZooKeeper and Hadoop files on central storage and extracting the desired ones from the checkpoint tar file.

IV. SECURITY

The primary security concern we had with these database systems were how they authenticated between the various components that run on one or more nodes of the system and communicated via the network. While different in the details, all the database systems handled this in similar ways: they have one set of credentials to authenticate other pieces of the system, where applicable, and a separate method to authenticate end-users of the system. We manage these two problems separately.

The database technologies contain many options for fine-grained access control. Accumulo in particular is known for its cell-level access control abilities. However the primary use case of MIT SuperCloud's database management system is analytics development in which these features are not commonly used. It would however be straightforward to extend the methods we use to address multiple user communities or individual authentication.

*A. Database components*

For authenticating between components of the system, the technologies utilize some form of shared secret. On Accumulo, this is unsurprisingly called the shared secret in the configuration files. For the SciDB and its underlying PostgreSQL, this is fundamentally the same, with the addition of a username. However, the username is simply to fit into the existing PostgreSQL security model and can be functionally ignored as there is only the SciDB user. The challenge then becomes securely distributing this secret value to all nodes and preventing unintended access to it.

We solve this problem by running the SGE prolog and epilog scripts described in Section III as a separate user on our system dedicated to running databases. The scripts will execute as this special user account regardless of which user requested the database start or stop. By then leveraging the Lustre shared file system available on all nodes, we ensure that only this special user can access the long-term shared secret or the configuration files that contain it.

The database technologies also come with a built-in superuser or root account. We manage the password for this account in the same manner.

*B. Database users*

MIT SuperCloud's security model is based on users' membership in a Linux security group. However passing through a user's current security context to any of these database systems is a considerable challenge. MIT SuperCloud enables users to execute analytic code from both nodes in the HPCC and their workstation, which may not be part of the same security infrastructure, which would be required for solutions such as Kerberos to function correctly. We therefore use the traditional password-based system that they all provide, albeit in a non-traditional way.

On each database we create a normal database user, with enough privileges to create tables and perform all other expected user operations. Each time the database is started, we generate a long random password to serve as an access key for this user account. This is stored on the Lustre shared file system available on all nodes and exported to user workstations as well, secured with standard Linux file permissions to permit only the users that are member of the database's associated security group. This password is changed in the database using the database superuser account on each startup.

Revocation of a user's permission to access the database is a two-step process: removal of their account from the associated Linux security group and restarting of the database to ensure the access key (password) is re-generated.

The D4M database bindings have been extended for MIT SuperCloud to automatically locate and provide the database access key during connection.

V. PEFORMANCE

The performance of these databases once running has already been well explored and is able to scale up to hundreds of nodes and 100,000,000 database inserts per second using Accumulo started through this system [Kepner 2014]. The primary performance metric of the MIT SuperCloud database management system is the speed at which it can perform the operations listed in Section III, in particular starting and stopping of the database. The critical step in both starting and stopping is the copy of the database data files from central storage on Lustre to local disk on startup and from local disk to Lustre on shutdown. The time taken by other steps involving in starting and stopping a database are small in comparison and do not vary significantly based on data size.

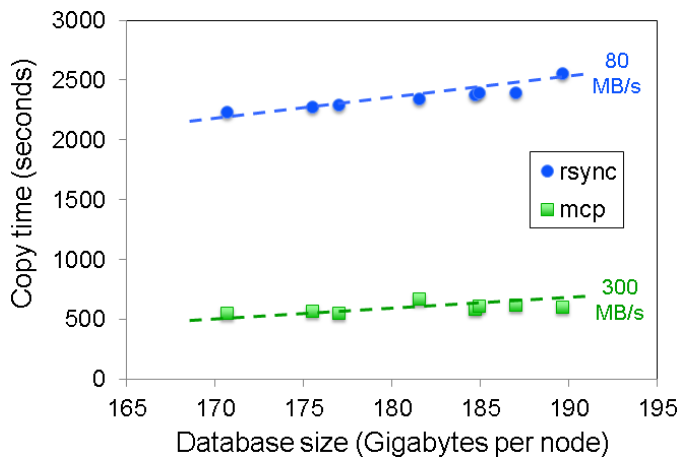

Figure 4. Copy time from central storage to local storage relative to database size per node, for an overall speed, for traditional rsync and multithreaded copy (mcp).

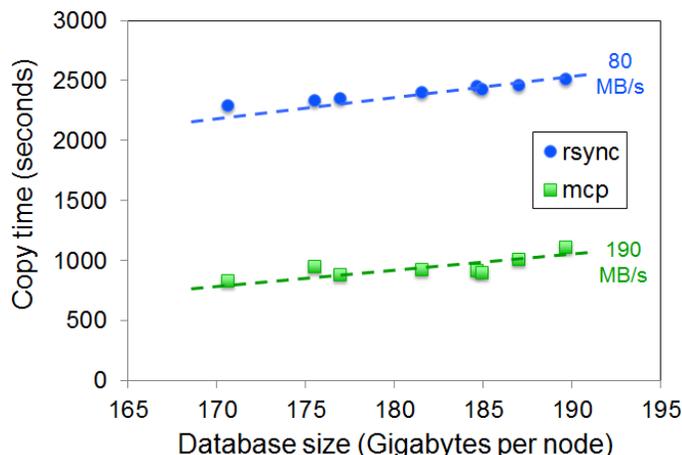

Figure 5. Copy time from local storage to central storage relative to database size per node, for an overall speed, for traditional rsync and multithreaded copy (mcp).

We have benchmarked two approaches to performing these copies: traditional single-threaded copying using rsync and a multi-threaded copy (mcp) using up to multiple concurrent instances of the "cp" command. We separately benchmarked the data flow in each direction: from central lustre storage to the local disk of each node, and from the local disk of each node back to central lustre storage.

As expected, the multi-threaded copy executing three operations in parallel takes full advantage of the Lustre parallel file system and significantly outperforms the single-threaded rsync. For the copy from central storage to local storage, this is shown in Figure 4. On an eight node system, we measured an aggregate bandwidth of 640 MB/sec for rsync compared to 2400 MB/sec for mcp.

The benefit, while still significant, is less on the copy from local storage to central storage. This is shown in Figure 5. On an eight node system, we measured an aggregate bandwidth of 640 MB/sec for rsync compared to 1520 MB/sec for mcp.

VI. SUMMARY

The MIT SuperCloud database management system allows for rapid creation and flexible execution of a variety of the latest scientific databases, including Apache Accumulo and SciDB. It has enabled us to run these databases on a HPCC as seamlessly as any other HPCC job. We are able to unbind the databases from any specific hardware and optimize the use of our computing hardware by allowing the databases to be stopped when not in use and started on-demand on any available hardware. We have also delegated the most common operations on these databases to the end user to be performed on-demand with no special system privileges required. By employing all these techniques, we have been able to significantly lower the barriers to using this advanced technology.